\documentclass[aps,twocolumn,pra,superscriptaddress,nofootinbib]{revtex4-2}
\pdfoutput=1

\usepackage{amsmath}
\usepackage{amssymb}
\usepackage{amsthm}
\usepackage{mathtools}
\usepackage{mathrsfs}
\usepackage{xfrac}
\usepackage{bbm}
\usepackage{bm, dsfont}
\usepackage[table]{xcolor}
\usepackage{graphicx}
\usepackage{natbib}
\usepackage[colorlinks=true,linkcolor=blue,citecolor=blue,urlcolor=blue]{hyperref}
\usepackage{cleveref}
\usepackage{verbatim, float}
\usepackage[normalem]{ulem}
\usepackage{physics,hyperref,nicefrac}	

\newtheorem{theorem}{Theorem}

\newtheorem*{result}{Result}

\newcommand{\be}{\begin{equation}}
\newcommand{\ee}{\end{equation}}

\newcommand{\cA}{\mathcal{A}}

\newcommand{\cG}{\mathcal{G}}

\newcommand{\id}{\mathds{1}}

\makeatletter
\newcommand{\undersim}[1]{\mathrel{\mathpalette\@undersim{#1}}}
\newcommand{\@undersim}[2]{%
  \vcenter{%
    \ialign{%
      ##\cr
      $\m@th#1#2$\cr
      \noalign{\nointerlineskip\kern.2ex}
      $\m@th#1\sim$\cr
      \noalign{\kern-.4ex}
    }%
  }%
}

\DeclareFontFamily{U}{mathx}{\hyphenchar\font45}
\DeclareFontShape{U}{mathx}{m}{n}{
      <5> <6> <7> <8> <9> <10>
      <10.95> <12> <14.4> <17.28> <20.74> <24.88>
      mathx10
      }{}
\DeclareSymbolFont{mathx}{U}{mathx}{m}{n}
\DeclareMathSymbol{\bigtimes}{1}{mathx}{"91}
\begin{document}

\title{Superactivation of genuine multipartite Bell nonlocality from two-party entanglement}

\author{Markus Miethlinger}
\author{Riccardo Castellano}
\author{Pavel Sekatski}
\author{Nicolas Brunner}
\affiliation{D\'{e}partement de Physique Appliqu\'{e}e,  Universit\'{e} de Gen\`{e}ve,  1211 Gen\`{e}ve,  Switzerland}

\begin{abstract}
    Characterizing the relation between entanglement and Bell nonlocality is a long-standing open problem, notably challenging in the multipartite case. Here, we investigate the effect of superactivation of genuine multipartite nonlocality. Specifically, we show that starting from multipartite states that feature only two-party entanglement (hence almost fully separable), it is possible to obtain GMNL in the many-copy regime. This represents the weakest possible resource for GMNL superactivation. On the technical side, we develop an efficient and practical criterion for certifying GMNL superactivation based on network entangled states, as well as a perfect parallel repetition result for the Khot-Vishnoi Bell game, which are of independent interest.
\end{abstract}

\maketitle 


\section{Introduction}

Quantum entanglement and Bell nonlocality have long been central to quantum research, from a foundational perspective but also for applications in quantum information processing. While entanglement arises at the level of the mathematical structure of quantum theory, Bell nonlocality is a property of measurement statistics, directly observable in an experiment.
Understanding the relation between entanglement and Bell nonlocality is one of oldest open problems in quantum information theory, which is still the focus of current research, see, e.g., \cite{BrunnerCavalcantiPironioScaraniWehner2014,Augusiak2014}.

It was first discovered by Werner \cite{Werner89}, and later generally proven by Barrett \cite{Barrett2002}, that certain entangled states---termed ``local''---admit a local hidden variable model, meaning that their statistics resulting from arbitrary local measurements can be exactly reproduced via a purely classical model. 
However, it was shown in turn that Bell nonlocality could nevertheless be activated for certain local entangled states by considering more elaborate Bell tests, involving, e.g., local filtering \cite{Popescu1995,Gisin1996,Hirsch2013}, local broadcasting \cite{Bowles2021,Boghiu2023,sekatski2025}, catalysis \cite{Bavaresco2025}, or in network scenarios \cite{Cavalcanti2011}. 

Perhaps the most striking instance of this effect is the superactivation of Bell nonlocality in the many-copy scenario, discovered by Palazuelos \cite{Palazuelos2012}. 
Starting from certain local entangled states $\rho$, it is possible to obtain Bell nonlocality by performing local (but joint) measurements on the state $\rho^{ \otimes k}$, where $k$ denotes the number of copies. 
Notably, this is possible for any entangled state with strong enough entanglement, as quantified by the entanglement fraction \cite{CavalcantiAcinBrunnerVertesi2013}. Other examples of many-copy activation have also been reported \cite{Liang2006,Navascues2011,Quintino2016}.

In the present work, we address these questions in the multipartite case, where both entanglement and nonlocality display a rich structure \cite{Bancal2013} and a complex relation \cite{Toth2006,Augusiak2015,Bowles2016}. 
Our focus is on the connection between entanglement and nonlocality in the many-copy scenario. 
In particular, we ask what are the minimal resources, in terms of entanglement, in order to obtain the strongest form of Bell nonlocality, namely genuine multipartite nonlocality (GMNL) \cite{Svetlichny1987}?

Previous works have shown that GMNL can be superactivated for strongly entangled multipartite states. First, it has been shown that any pure genuine multipartite entangled (GME) state becomes GMNL in the many-copy regime (while this is still open in the single-copy regime) \cite{ContrerasTejadaPalazuelosdeVicente2021}. Second, it is proven that certain mixed states that are GME but not GMNL (as they admit a biseparable local model) become GMNL for a sufficient number of copies \cite{ContrerasTejadaPalazuelosdeVicente2022}. In parallel, it has been shown that biseparable (i.e. non-
GME) states become GME in the many-copy regime \cite{Yamasaki2022,Palazuelos2022,Weinbrenner2025}.

Here we show that GMNL can in fact be superactivated starting from much weaker resources, namely states that feature only two-party entanglement. 
More precisely, we construct $N$-partite entangled states $\rho$ that are $(N-1)$ separable, i.e., almost fully separable. 
Then we show that $\rho^{\otimes k}$ becomes GMNL for some large enough $k$. Hence, we have identified the weakest possible entanglement resource for superactivating GMNL. On the more technical level, our proof relies on new results on multipartite extensions of bipartite Bell games, and a perfect parallel repetition result for the Khot-Vishnoi (KV) game \cite{KhotVishnoi2005,BuhrmannRegevScarpadeWolf2011}, which are of independent interest.


\section{Setting the scene}\label{sec:Preliminaries}

Consider a set of $N$ distant parties sharing a multipartite entangled state $\rho$. 
Each party $\mathcal{A}_i$, upon receiving a classical input $x_i$, performs a measurement, represented by a POVM $\{M_{a_i|x_i}\}_{a_i}$, and produces a classical output $a_i$. 
The resulting probability distribution is given by 
\begin{align}\label{eq:bornrule}
    &P(\bm a|\bm x)=\Tr(M_{a_1|x_1}\otimes \ldots \otimes M_{a_N|x_N} \rho ), 
\end{align}
where $\bm a \coloneqq \bigtimes_{i}a_i$ and $\bm x \coloneqq \bigtimes_{i}x_i,$.
Such a distribution can exhibit different forms of multipartite Bell nonlocality, the strongest one being GMNL. 
Essentially, one asks here whether the observed distribution could be explained by a classical model, where the parties are merged into two subgroups such that all parties in each subgroup can jointly produce their outcomes based on all inputs. 
Additionally, probabilistic mixtures over different subgroups are also taken into account. 
Such “biseparable” local models take the following form
\begin{align}
    \begin{split}\label{eq:bilocal}
        &P(\bm a|\bm x)=\sum_\lambda \mu(\lambda) P_{bp}(\bm a_\lambda,\overline{\bm a_\lambda}|\bm x_\lambda, \overline{\bm x_\lambda},\lambda),
    \end{split}\\
    \begin{split}
        &P_{bp}(\bm a_\lambda,\overline{\bm a_\lambda}|\bm x_\lambda, \overline{\bm x_\lambda}, \lambda) =P_{\mathcal{G}(\lambda)}(\bm a_\lambda |\bm x_\lambda,\lambda) \\
        &\hspace{94pt}\cdot \hspace{2pt}\hspace{1pt} P_{\overline{\mathcal{G}(\lambda)}}(\overline{\bm a_\lambda} |\overline{\bm x_\lambda},\lambda),
    \end{split}\label{eq:biproductbehavior}
\end{align}
where we refer to $P_{bp}(\bm a_\lambda,\overline{\bm a_\lambda}|\bm x_\lambda, \overline{\bm x_\lambda}, \lambda)$ as biproduct behaviors, over the bipartition $\mathcal{G}(\lambda)|\overline{\mathcal{G}(\lambda)}$, and $\bm x=\bm x_\lambda \times \overline{\bm x_\lambda}$, $\bm a=\bm a_\lambda \times \overline{\bm a_\lambda}$ are the inputs and outputs of all parties, respectively, and $\bm{a}_\lambda, \overline{ \bm{a}_\lambda}$ and $\bm{x}_\lambda, \overline{\bm{x}_\lambda}$ collect the inputs and outputs for $\mathcal{G}(\lambda),\overline{\mathcal{G}(\lambda)}$, respectively. 

A distribution for which a decomposition of the above form does not exist exhibits GMNL. 
While several definitions have been proposed \cite{Bancal2013}, depending on the constraints imposed on the distributions $P_{\mathcal{G}(\lambda)}(\bm a_\lambda |\bm x_\lambda,\lambda)$ in Eq.~\eqref{eq:bilocal}, we consider here that no special restrictions are imposed on these distributions, as in the original approach of Ref. \cite{Svetlichny1987}. This leads to the strongest notion of GMNL. 
Such distributions can be obtained via strongly entangled states, such as GHZ states \cite{MitchellPopescuRoberts2004} and Dicke states \cite{WangMarkham2012}. 
In particular, the observation of GMNL implies that the underlying state $\rho$ features GME \cite{HorodeckiRudnickiZyczkowski2025}, see, e.g., \cite{JiaZhaiYuWuGuo2020}. 

The problem of interest in the present work is in fact the inverse question. 
What are the minimal resources, in terms of entanglement, that can give rise to GMNL? 
In particular, we want to address this question in the multi-copy Bell scenario, i.e., where the underlying state consists of a finite number of copies of a given resource state $\sigma$. 
This means that in Eq.~\eqref{eq:bornrule}, we now consider states of the form $\rho = \sigma ^{\otimes k}$, where $k$ denotes the number of copies of the resource state $\sigma$. 
In such a scenario, each party performs a joint quantum measurement on $k$ local subsystems. 
As these measurements can be performed in some non-separable basis of the $k$ subsystems, this can lead to the phenomenon of superactivation of Bell nonlocality, first investigated in the bipartite scenario \cite{Palazuelos2012,CavalcantiAcinBrunnerVertesi2013} and more recently in the multipartite scenario \cite{ContrerasTejadaPalazuelosdeVicente2021,ContrerasTejadaPalazuelosdeVicente2022}. 
Loosely speaking, this means that starting from a resource state $\sigma$ which is weakly entangled, it is possible to obtain strong nonlocality in the multi-copy regime. 

Below we present a construction for identifying the minimal resources for superactivation of GMNL. 
Specifically, we prove that starting from a resource state that has only two-party entanglement, hence is almost fully separable---formally $(N-1)$-separable \cite{Horodeckitimes42009}---it is possible to superactivate GMNL. This shows that, in the many-copy regime, the strongest form of multipartite Bell nonlocality, i.e., GMNL, can be obtained starting from the weakest possible form of multipartite entanglement.

Additionally, our result shows that GMNL superactivation is also possible starting from resource states that are (at most) only weakly Bell nonlocal. Specifically, our states are $(N-1)$-local, i.e., there always exists a decomposition of the measurement statistics into $(N-1)$-product behaviors, in analogy to Eq.~\eqref{eq:biproductbehavior}. Finally, we discuss the question of the superactivation of GMNL from a state that admits a fully local ($N$-local) model.

\section{Superactivation of GMNL}
\begin{figure}
    \centering
    \includegraphics[width=0.95\linewidth]{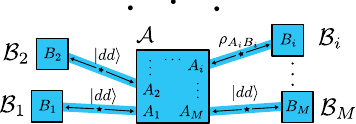}
    \caption{Structure of the state for GMNL superactivation. In each round of state preparation in the network, a shared random variable $\lambda = i$ determines the preparation of the state $\sigma_i = \rho_{A_iB_i}\bigotimes_{j\neq i}\ketbra{dd}_{A_jB_j}$. This state consists of a bipartite entangled state shared between Alice $\mathcal{A}$ and the $i$-th Bob $\mathcal{B}_i$, while the remaining Bobs receive a locally orthogonal flag state $|dd\rangle$ shared with Alice.}
    \label{fig:starnetwork}
\end{figure}

We consider multipartite entangled states created on a quantum network, where a number of sources distribute bipartite entangled states to the parties (nodes).  For clarity, we focus on the case of a star network with $N=M+1$ parties, see Fig.~\ref{fig:starnetwork}. The central party Alice $\mathcal{A}$ holds $M$ subsystems $\{A_{i}\}_{i=1}^M$, while the outer parties $\mathcal{B}_i$ receive one subsystem $B_i$ each. We consider particular states of the form
\begin{align}
    \sigma^\star&\coloneqq \frac{1}{M}\sum_{i=1}^M \sigma_i \label{eq:thestate}\coloneqq \frac{1}{M}\sum_{i=1}^M 
\big(\rho_{A_i B_i} \bigotimes_{j\ne i} |dd\rangle\!\langle dd|_{A_j B_j}\big).
\end{align}
The state $ \sigma^\star$ is a mixture of $M$ terms. 
For each term, only one of the sources (on the link $A_{i}B_{i}$) produces a bipartite entangled state $\rho_{A_i B_i}$, while all other sources distribute a classically correlated (flag) state, in an orthogonal degree of freedom, i.e., $\ket{dd}$. 

Clearly, such a state has only weak multipartite entanglement, as it is constructed using only two-party entangled states. 
It is $M$-separable, hence close to being fully separable and far from being GME.
Nevertheless, we will show that such a state can exhibit extremely strong correlations; taking sufficiently copies of $ \sigma^\star$ we obtain GMNL, the strongest form of multipartite nonlocality. 

A crucial figure of merit for our result is the entanglement (or singlet) fraction $F_i$ \cite{HorodeckiHorodecki1999}, defined as 
\begin{align}
    F_i \coloneqq \max_{\Phi} \Tr(\Phi\rho_{A_i B_i}),
\end{align}
where one maximizes over all maximally entangled states $\Phi^{}\coloneqq |\Phi^{}\rangle\!\langle\Phi^{}|$. We now show our main result.

\begin{result}\label{thm:mainresult}
    For a sufficiently large number of copies $k$, the state $(\sigma^{\star})^{ \otimes k}$ becomes GMNL if $\prod_{i=1}^M F_i>d^{-1}$.
\end{result}

Let us sketch the proof of the result. The first step is to observe that since the correlated flag states $|dd\rangle$ are locally orthogonal to $\rho$, each party can determine if it received entanglement from the corresponding source with the local non-demolition measurement $\{\mathbbm{1}_{d^2-1}, |d\rangle\!\langle d|\}$. 
Hence, given access to sufficiently many copies of the state $\sigma^\star$ and using local operations, the parties can prepare the state $\rho_{\mathcal{A}\mathcal{B}_1\dots\mathcal{B}_M}=\bigotimes_{i=1}^M \rho_{A_i B_i}$ satisfying
\begin{align}
    \Tr \left[ \rho_{\mathcal{A}\mathcal{B}_1\dots\mathcal{B}_M} \bigotimes_{i=1}^M \Phi^+_{A_i B_i}\right]>\frac{1}{d} \label{eq: >1/d}
\end{align} 
with any desired probability $p<1$. 
Given even more copies of $\sigma^\star$, they can prepare any number of states $\rho_{\mathcal{A}\mathcal{B}_1\dots\mathcal{B}_M}^{\otimes k}$ with arbitrarily large probability. 
For large enough $k$, this state is GMNL as follows from Theorem~\ref{thm:entanglementfraction}---the premise of the Theorem (Eq.~\ref{eq:thmentanglementfraction}) becomes Eq.~\eqref{eq: >1/d} for the star network. 

The next section is devoted to the proof of the Theorem, which we give for a general network. 
To do the proof we first demonstrate two preliminary results. 
Theorem~\ref{thm:repetitionsVSlocal} presents a lifting of a bipartite Bell game $\mathrm{G}$ to a multipartite network. 
Using the resulting network game to witness GMNL requires a local bound for {\it parallel repetitions} of $\mathrm{G}$. 
For this purpose, the following Theorem~\ref{thm:ppr} presents such a bound for a specific Bell game.

\section{Methods}

In this section, we present methods for demonstrating GMNL (and its superactivation) that are tailored to the network configuration. 
These methods are versatile and apply to arbitrary networks with bipartite sources. Specifically, we devise Bell inequalities that naturally accommodate an arbitrary number of parties and various connectivities of these parties. 
Following \cite{ContrerasTejadaPalazuelosdeVicente2021}, we consider multipartite Bell games that are constructed out of bipartite ones. 
We provide a number of results for the characterizing of such network extensions of bipartite Bell games, that can be of independent interest.

Specifically, consider a bipartite Bell game ${\rm G}$ with a score given by 
\begin{equation}\label{eq:bellgame}
    {S} = \sum_{a,b,x,y}  { G}[a,b,x,y] P(a,b|x,y) p(x,y), 
\end{equation}
where $P(a,b|x,y)$ is the probability that Alice and Bob produce the outputs $a,b$ given the inputs $x,y$, ${G}[a,b,x,y] \in [0,1]$ is the winning condition, and $p(x,y)$ is the distribution of inputs $x,y$. 
The bound $S_L^\mathrm{G}$ denotes the local bound, i.e., the maximal score achievable for local models, i.e., where $P(\bm a | \bm x)$ is of the form of Eq.~\eqref{eq:bilocal} in the case of $N=2$ parties. In the following, we omit the superscript $\mathrm{G}$, when there is no ambiguity on which game $\mathrm{G}$ we consider the local score for, simply writing $S_L$ instead.

\begin{figure}[t]
    \centering
    \includegraphics[width=.75\linewidth]{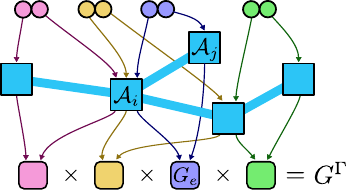}
    \caption{Network-extension game ${\rm G}^\Gamma$ of a bipartite Bell game $\rm G$. All pairs of parties $\mathcal{A}_i\mathcal{A}_j$ connected by the network graph $\Gamma_{ij}=1$ (blue) play the bipartite game $G_{e}$, with $e=\{i,j\}$, receiving inputs and providing outputs (shown in pink, yellow, purple and green). The $G_e$ of all bipartite games are multiplied to get the winning condition of the global game ${G}^\Gamma=\prod_{e\in E} G_e$.}
    \label{fig:network}
\end{figure}

We can now define the network extension of the bipartite game as follows. Consider a network with $N$ parties $\mathcal{A}_i$ connected by a graph with the adjacency matrix $(\Gamma_{ij})$. In the network extension ${\rm G}^{\Gamma}$ of the game ${\rm G}$ (see Fig.~\ref{fig:network}), an instance of the game ${\rm G}$ is played between all pairs of connected parties, i.e., all $\cA_i$,$\cA_j$ with $\Gamma_{ij}=1$.
The associated score $S^\Gamma$ reads
\begin{align}
    {S}^\Gamma \!\!=&\! \sum_{\bm a,\bm x}\!  P(\bm a|\bm x)  
    \!\prod_{\substack{
    e\in E
    }}\!
    { G}[a_e^{(i)},a_e^{(j)},x_e^{(i)},x_e^{(j)}] \,p(x_e^{(i)}, x_e^{(j)}),\label{eq:networkscore}\\
    \bm a \coloneqq& \bigtimes_{e \in E}(a_e^{(i)},a_e^{(j)}),\quad 
    \bm x \coloneqq \bigtimes_{e \in E}(x_e^{(i)},x_e^{(j)}),
\end{align}
where $E\coloneqq\{\{i,j\}|\Gamma_{ij}=1\}$. 

A key question is now to compute a lower bound on the score ${S}^\Gamma$ which would certify GMNL, i.e., guarantee that the observed statistics does not admit a biseparable local models of the form of Eq.~\eqref{eq:bilocal}. 
It turns out that this question is intimately connected to a property of the bipartite game ${\rm G}$ known as \emph{parallel repetition} \cite{Raz1998}. 
Specifically, consider that the parties ($A$ and $B$) play $k$ games ${\rm G}$ in parallel. 
That is, in each round of the game, they receive $k$ inputs $x_i$ and $y_i$ (sampled independently with $p(x_i,y_i)$) and report $k$ outputs $a_i$ and $b_i$ used to compute the product score $\Pi_{i=1}^{k} {G}[a_i,b_i,x_i,y_i]$. 
Its expected value, denoted $S^{\otimes k}$, is the score of the game ${\rm G}^{\otimes k}$, referred to as the $k$-repetition of ${\rm G}$.
The players may of course play the $k$ instances of the game entirely independently, resulting in a score $S^{\otimes k} = S^{k}$, but there can exist correlated strategies that lead to higher scores \cite{BarrettCollinsHardyKentPopescu2002}.

In general, we obtain the following criteria for certifying GMNL, applicable to an arbitrary network and for the corresponding network extension of any bipartite game $\rm G$. 

\begin{theorem}\label{thm:repetitionsVSlocal}
    Consider a behavior $P(\bm a|\bm x)$ and a network extension game $\rm{G}^\Gamma$ of the bipartite game $\rm{G}$ given the network graph $\Gamma$. If the score $S^\Gamma$ Eq.~\eqref{eq:networkscore} for this behavior fulfills ${S}^\Gamma > {{S}^{\otimes c}_L}$---where $c$ is the capacity of the min-cut of the graph $\Gamma$---then $P(\bm a|\bm x)$ is GMNL. %
\end{theorem}

Note that here we implicitly assumed that the game ${\rm G}$ is symmetric under the permutation of the players; To extend the result to asymmetric games, one can consider directed graphs.
For the complete proof see Appendix~\ref{app:proofrepetitionsVSlocal}, but to give a brief idea of the proof, consider the following.
Any behavior that is not GMNL can be expressed as a sum of biproduct behaviors Eq.~\eqref{eq:bilocal}, where each bipartition $\mathcal{G}(\lambda)|\overline{\mathcal{G}(\lambda)}$ is taken over the parties of the network.
Since the marginal probability distributions are unrestricted, one can marginalize over all instances of the game that are played within the groups $\mathcal{G}(\lambda)$ and $\overline{\mathcal{G}(\lambda)}$, while the games between the groups are still played using local resources.
Then, an upper bound for the score of the local strategy is ${{S}^{\otimes c_\lambda}_L}$, where $c_\lambda$ is the capacity of the cut $\mathcal{G}(\lambda)|\overline{\mathcal{G}(\lambda)}$ of $\Gamma$.
The strongest bound is then achieved for a biproduct behavior for which the bipartitions $\mathcal{G}(\lambda)|\overline{\mathcal{G}(\lambda)}$ are min-cuts of the network. 
Thus, for biseparable local models, the bound of the network extension game $\rm{G}^\Gamma$ is given by the local bound of the game $\rm{G}^{\otimes c}$, i.e., the $c$-repetition of $\rm{G}$ played over a bipartition of the parties in the network.

This permits the investigation of GMNL superactivation for a variety of games and networks. 
However, to evaluate the bounds provided by Theorem~\ref{thm:repetitionsVSlocal}, one needs upper bounds on the local bound of $k$-repetitions of bipartite games. 
These bounds depend on the specific game considered, with some games providing stronger bounds on the local score than others, see e.g. \cite{BrunnerCavalcantiPironioScaraniWehner2014}.

For our problem, the most relevant game is the Khot-Vishnoi (KV) Bell game \cite{KhotVishnoi2005}, for which strong bounds have been derived \cite{BuhrmannRegevScarpadeWolf2011}. 
This game is also the main tool for demonstrating superactivation of Bell nonlocality in the bipartite case \cite{Palazuelos2012,CavalcantiAcinBrunnerVertesi2013}. 
Its extension to star networks has been discussed in \cite{ContrerasTejadaPalazuelosdeVicente2022}. 
Here, we provide more general and stronger bounds by showing that the KV game obeys \emph{perfect parallel repetition}, making it an ideal Bell game for our purpose.

\begin{theorem}\label{thm:ppr}
    Given a $k$-repetition of the KV game, the local bound $S_L^{\otimes k}$ is given by
    \begin{align}
        S_L^{\otimes k}= (S_L)^k 
    \end{align}
    where $S_L$ is the local bound for a single instance of the KV game, in the asymptotic limit of infinitely large inputs.  
\end{theorem}

The proof closely follows the derivation of the local bound for the KV game presented in Ref.~\cite{BuhrmannRegevScarpadeWolf2011}, and is given in  Appendix~\ref{app:parallelrepetitionbounds}.

Finally, combining Theorems \ref{thm:repetitionsVSlocal}
and \ref{thm:ppr} leads to the following Theorem, which provides a general criterion for GMNL superactivation. Applied to the special case of the star network, this establishes our main result announced earlier.
\begin{theorem}\label{thm:entanglementfraction}
    Consider a multipartite state $\rho$ distributed to parties ${\mathcal{A}_1,\ldots,\mathcal{A}_N}$, over a network with adjacency matrix $\Gamma$. Then $\rho$ is many-copy-GMNL if
    \begin{align}\label{eq:thmentanglementfraction}
        F^\Gamma \coloneqq \langle \Phi^{\Gamma}| \rho |\Phi^{\Gamma}\rangle > \frac{1}{d^c},
    \end{align}
    where $|\Phi^{\Gamma}\rangle\coloneqq\bigotimes_{e\in E} |\Phi^{+}\rangle_{A_{i}^e A_{j}^e}$, and $c$ is the capacity of the min-cut of the graph specified by $\Gamma$.
\end{theorem}
Note that, in the definition of $\Phi^{\Gamma}$, there is an implicit optimisation over the choice of the maximally entangled states $|\Phi^{+}\rangle_{A_{i}^e A_{j}^e}$.
More importantly, note that the criterion applies to general states, i.e. $\rho$ does not need to factorizes like $|\Phi^{\Gamma}\rangle$ with respect to the network. See Appendix~\ref{app:GMNLactivationlemma} for the proof, which we outline in the following.
The key idea is to note that the (isotropic?) twirling channel $\mathcal{T}[{\cdot}]\coloneqq \int \dd U  (U\otimes U^*) {\cdot}  (U\otimes U^*)^\dag$ can also be expressed in the following way \cite{Rains2001}
\begin{align}
    \mathcal{T}[{\cdot}] = \Phi^+ \Tr(\Phi^+\cdot) + \Tr\left[(\id-\Phi^+)\cdot\right] (\id-\Phi^+).
\end{align}
Hence, applied on all pairs of systems connected by a source, we get
\begin{align}
    \widetilde \rho =\bigotimes_{e\in E}\mathcal{T}_{A_{i}^e A_{j}^e} [\rho] = F^\Gamma\,  \Phi^\Gamma + (1-F^\Gamma)\, \tau
\end{align}
where $\tau\succeq 0$. 
When $F^\Gamma>\frac{1}{d}$, playing the network extended ${\rm KV}^\Gamma$ of the bipartite ${\rm KV}_{d^k}$ game with $k$ copies of this state $\widetilde \rho^{\otimes k}$, the parties can achieve a global score ${F^\Gamma}^k S^{\Gamma}_{\mathrm{KV}}({\Phi^\Gamma}^{\otimes k})>S_{\mathrm{KV},L}$, for large enough $k$.
Then $\rho$ is GMNL by Theorem~\ref{thm:repetitionsVSlocal}.

\section{GMNL from Fully Local state?}
A related, but different question is whether GMNL can superactivated starting from an $N$-partite multipartite entangled state that admits a fully local model, i.e. that the statistics arising from any possible local measurements on $\rho$, as in \eqref{eq:bornrule}, can always be reproduced by a model of the form 
\begin{align}
    \label{eq:fullyLocal}
        &P(\bm a|\bm x)=\int d\lambda \mu(\lambda) \prod_{i_1}^N{P(a_i|x_i \lambda)} \,.
\end{align}

Our construction for the main result comes close to this. 
Clearly, the state in Eq.~\eqref{eq:thestate} admits a local model where two parties join, while the $N-2$ others are all separated. To see this, simply consider again that the state $\sigma_i$ in Eq.~\eqref{eq:thestate} is a $(N-1)$-product state, since only two parties share nonlocal correlations. Therefore, by convexity, $\sigma^\star$ is $(N-1)$-local as well.

A promising alternative is to move to a different network, where Theorem~\ref{thm:entanglementfraction} provides the strongest bound on superactivation.
This is achieved for fully connected networks, which maximize the capacity of the min-cut $c$, given a number of parties $N$.
Moreover, the strongest bound is achieved for $N=3$, the triangle network which represents the minimal non-trivial example. Specifically, consider a triangle network with three parties $A$, $B$ and $C$, each holding two subsystems. Consider the following tripartite state 
\begin{align}
\begin{split}
    \sigma^{\Delta}(F)= \tfrac{1}{3}(&\rho(F)_{A_2B_1}\otimes \ketbra{2222}_{B_2 C_1 C_2 A_1} \\
    +& \rho(F)_{B_2C_1}\otimes\ketbra{2222}_{C_2 A_1 A_2 B_1} \\
    +&\rho(F)_{C_2A_1}\otimes\ketbra{2222}_{A_2 B_1 B_2 C_1}).
\end{split}
\end{align}
where $\rho(F) =  \Phi^{+} + \frac{1-F}{3}(\mathbbm{1}_{4}-\Phi^{+})$ is a two-qubit isotropic state. Now observe, that the global state $\sigma^{\Delta}(F)$ would admit a fully local model if $\rho(F)$ admits a local model. 
This is the case for low enough values of the entanglement fraction $F$. A subtle, but important point there is that the local model for $\rho(F)$ should cover the most general local measurements, namely POVMs. 
This is because each party holds two subsystems, hence they can in principle use the auxilliary systems (in the flag state $\ket{2}$) in order to realize arbitrary POVMS on the qubit space $\{\ket{0},\ket{1}\}$ via a projective measurements on the qutrit space. 
While local models for POVMs exist for $\rho(F)$ they (unfortunately) work only for much lower parameters values of $F$. Specifically, the best known bound is $F\le 0.625$ 
\cite{ZhangChitambar2024,Renner2024}, whereas our Theorem~\ref{thm:entanglementfraction} requires $F>{2^{-\frac{2}{3}}}\approx 0.63$. 

We are confident that this small gap could be closed by deriving better local models for POVMs. 
First, the models in \cite{ZhangChitambar2024,Renner2024} are of a specific form (namely a local hidden state (LHS) model \cite{wiseman2007}) and hence one could expect stronger local models to exist (not of LHS form). 
Indeed, much stronger local models are known for the case of projective measurements, achieving $F_L^{\rm PM}\approx 0.762$ \cite{HirschQuintinoVertesiNavascuesBrunner2017,DesignolleVertesiPokutta2026}. An alternative approach could investigate whether GME states admitting a fully local model \cite{Bowles2016} could lead to GMNL in the many-copy regime. The challenge here is to find intrinsic multipartite Bell games with an unbounded violation \cite{PalazuelosVidick2016}. 

\section{Discussion and Outlook}\label{sec:Discussion}

We demonstrated that GMNL superactivation is possible starting from multipartite states that feature only two-party entanglement. 
From the perspective of entanglement, this represents the weakest possible resources for reaching GMNL. 
This question is also interesting from the perspective of nonlocality. Here we have proven that GMNL can be superactivated from states that amit an ($N-1$)-local model. 
The weakest possible resource would be an $N$-partite entangled state admitting a fully local (or $N$-local) model, which it would be interesting to address.

Our work also brings a number of technical tools for the investigation of GMNL that could find broader applications. 
In particular, Theorem~\ref{thm:repetitionsVSlocal} provides a natural framework by lifting bipartite Bell inequalities to multipartite ones for GMNL. 
It also naturally connects to the notion of parallel repetition, exhibiting its relevance in the context of GMNL. 
Moreover, Theorem~\ref{thm:ppr}, establishing asymptotic perfect parallel repetition for the KV game, is of independent interest. 
Indeed the KV game provides an explicit example a Bell game with unbounded violation (with nearly optimal quantum to classical gap), useful in many different contexts \cite{PalazuelosVidick2016}.

Finally, Theorem~\ref{thm:entanglementfraction} provides a general and practical criterion for GMNL superactivation that can be used for broad classes of states and general networks. 
This offers a useful tool for identifying new instances of superactivation and for further exploring the relationship between entanglement and nonlocality in multipartite systems.

\section{Acknowledgments}

We thank Tamás Vértesi, Jessica Bavaresco, and Bora Ulu for helpful discussions, and acknowledge financial support from the Swiss National Science Foundations (project 219366 and NCCR-SwissMAP).


\bibliographystyle{apsrev4-1fixed_with_article_titles_full_names_new}
\bibliography{refs}

\appendix
\section{Proof for Theorem~1}\label{app:proofrepetitionsVSlocal}
By definition, a behavior that is not GMNL is a mixture of biproduct behaviors, see Eqs.~\eqref{eq:bilocal} and~\eqref{eq:biproductbehavior} in the main text. 
By linearity, the total score for such behaviors is the average over $\lambda$.
Therefore, we now bound the score for a fixed $\lambda$ and define 
\begin{align}
    P_{bp}(\bm a,\overline{\bm a}|\bm x, \overline{\bm x})\coloneqq P_{bp}(\bm a_\lambda,\overline{\bm a_\lambda}|\bm x_\lambda, \overline{\bm x_\lambda}, \lambda),
\end{align}
dropping the subscripts. 
Then, the biproduct behavior $P_{bp}(\bm a,\overline{\bm a}|\bm x, \overline{\bm x})$ is local with respect to the partition $\cG|\overline{\cG}$, where $\cG\coloneqq \cG(\lambda)$.
Now, we denote the random variables $G_{(i,j)} \coloneqq {\rm G}[a_i,a_j,x_i,x_j]$, i.e., where $G_{(i,j)}$ corresponds to the game being played between the pair $(\mathcal{P}_i,\mathcal{P}_j)$.
Using this, we define some relabeling $\{G_k\}_k=\{G_{(i,j)}\}_{(i,j)\in\gamma}$, such that $G_k=G_{(i,j)}$, for some given $k$ and $(i,j)$.
Without loss of generality, we assume that $G_k$ with $k\in g\coloneqq\{1,\ldots,c\}$ corresponds to a game $G_{(i,j)}$ between parties $\mathcal{P}_i\in\cG$ and $\mathcal{P}_j\in\overline{\cG}$, i.e., the games that are played between parties that are partitioned by $\cG|\overline{\cG}$.\\
Then, because the partition $\cG|\overline{\cG}$ induces some cut $C_\mathcal{G}$ of the graph of $N$, the games $G_k$ with $k\in g$ correspond to edges of the cut-set $X_\mathcal{G}$ of $C_\mathcal{G}$, i.e., the edges that connect a party $\mathcal{P}_i\in\cG$ to a party $\mathcal{P}_j\in\overline{\cG}$.
Therefore, the number of games being played between the groups $c$ is equal to the size of the cut-set $|X_\mathcal{G}|$, which for unweighted graphs is equal to the capacity.
The total score for the multipartite game is then
\begin{align}
\begin{split}
    {S}_\Gamma &= \mathds{E}[G_1\cdot G_2\cdots G_{|\gamma|}].
\end{split}
\end{align}
We can find an upper bound to this expression, by considering that $0\leq G_k\leq 1$, that $P_{\mathcal{G}}(a_k|x_k),P_{\overline{\mathcal{G}}}(\overline{a_k}|\overline{x_k})$ are any signaling distributions, and that the marginal distribution
\begin{align}
\begin{split}\label{eq:localmarginal}
    P_{bp}(\bm a_g, \overline{\bm a_g}|\bm x_g,\overline{\bm x_g}) \coloneqq \!\!\!\!\!\!
    \sum_{a_i,\overline{a_i},x_i,\overline{x_i}|i\notin g} \!\!\!\!\!\!
    P_{bp}(\bm a,\overline{\bm a}|\bm x, \overline{\bm x})
    \cdot\prod_{i\notin g}p_i(x_i, \overline{x_i})
\end{split}
\end{align}
is manifestly product. 
Then, we find that each game $G_k$, where $k\notin g$, can be won with unit probability, and only the games $k\in g$ remain, which only have access to local resources, because Eq.~\eqref{eq:localmarginal} is local with respect to $\cG|\overline{\cG}$.
Following from this, we have the following upper bound
\begin{align}
\begin{split}
    \mathds{E}[G_1\cdot G_2\cdots G_{|\gamma|}]
    \leq\mathds{E}[G_1 \cdots G_c]=S^{\otimes c}_L.
\end{split}
\end{align}
Hence, we find for fixed $\lambda$ that
\be
 { S}_\Gamma \leq  S^{\otimes c}_L.
\ee
Since this holds for all $\lambda$ we find that by convexity
\be
P(\bm a|\bm x) \text{ not GMNL } \implies { S}_\Gamma \leq  S^{\otimes c}_L,
\ee
concluding the proof. 


\section{Bounds on Local Strategies for Parallel Repetitions of KV}\label{app:parallelrepetitionbounds}
\subsection{k-Repetitions of KV}
We first give a brief description of parallel repetitions of the KV game between two parties, and then we prove an upper bound to the winning probability that can be achieved with local strategies. 
The reader unfamiliar with the KV game is suggested to assume $L=1$ at first read, as this corresponds to the standard KV game. \\
We use $+$ for the sum mod $2$ of bit strings and $\times$ for the direct sum, which is equivalent to joining the strings. 
Equipped with the mod $2$ addition the set of $n$-digit bit-string, $(\{0,1\}^{n},+)$ is a group of order $2^n$. For $n:=2^{k}$ the Hadamard codewords  
\begin{equation}
H_{n}:=\{h: \exists \,a \in \{0,1\}^{k}\;\; \text{s.t.}\; h_{j}= a \cdot j \quad \forall j \in \{0,1\}^{k}\}
\end{equation}
form a subgroup of $(\{0,1\}^{n},+)$, with $2^k$ elements labeled by $a$. The Hadamard code has been studied in classical communication theory and computer science for its application as a linear error correction code \cite{vanLint1999}. Indeed it can be proved that the Hamming weight of each bit-string (except for the $h=\bm{0}$) equals $\abs{h}:=\sum_{i=0}^{n-1}h_{i}=\frac{n}{2}$. This implies that the Hamming distance between any distinct pair of bit-strings in $H_{n}$ equals $\frac{n}{2}$, indeed 
\begin{equation}
    \forall \; h_{1}\neq h_{2} \in H_{n}\quad \abs{h_{1}-h_{2}}=\abs{h_{1}+h_{2}}=\frac{n}{2}
\end{equation}
Furthermore, $H_{n}$ yields the subgroup of $(\{0,1\}^{n},+)$ with the largest possible order \cite{MacWilliamsSloane1988}. This very property of the Hadamard code is at the heart of the near optimality of the KV game.  

We consider two parties, Alice and Bob.
Now, Alice receives $L$ inputs $x_{1},\ldots, x_{L} \in \{0,1\}^{n}$, where we denote the joint input with $\bm x= \underset{i=1}{\overset{L}{\bigtimes}}x_{i}$.
The joint input $\bm x$ of Alice is randomly sampled from a uniform probability distribution over $\{0,1\}^{n}$, i.e., $P(\bm x)=2^{-n}\  \forall \bm{x}$. 
Then, Bob receives a joint input $\bm y:=\bm x + z$, where $z$ is a $Ln$-digit bit string, where each bit was sampled independently from a biased probability distribution $p(1):=\eta \in (0,1/2)$. \\
The outputs $a(\bm x)$ and $b(\bm y)$ that Alice and Bob, respectively, give, must be contained in the orbit of their inputs under the action of the $L$-times cartesian product of the Hadamard group, which we denote by $H_{n}^{L}:=\{\underset{i=1}{\overset{L}{\times}}h_{i}: h_{i}\in H_{n}\}$, i.e.,
\begin{align}
  &a(\bm x) \in O_{H}(\bm x):=\{\bm x + h: h\in H_{n}^{L}\} , \\
  &b(\bm y)\in O_{H}(\bm y).
\end{align}
Further, their outputs have to be constant over the orbits
\begin{align}
& a(\bm x)=a(\bm x+h) \quad \forall h\in H_{n}^{L}, \\
&  b(\bm y)=b(\bm y+h) \quad \forall h\in H_{n}^{L}.
\end{align}
Equivalently, their inputs can be seen as $O_{H}(\bm x), O_{H}(\bm y)$ instead of $\bm x,\bm y$. For this, recall that the group orbits are all disconnected, i.e., we have either $O_{H}(\bm x)=O_{H}(\bm x')$ or $O_{H}(\bm x)\cap O_{H}(\bm x')= \emptyset$.
They win the game if and only if $a(\bm x)+b(\bm y)=z$.

\subsection{Preliminaries}
To complete the proof, some basic properties of real functions on the boolean cube have to be reviewed, see, e.g., \cite{BuhrmannRegevScarpadeWolf2011} for more details. 
Let $\mathcal{L}_{n}:=\{F: \{0,1\}^{n}\rightarrow \mathbb{R}\}$ be the real vector space of real functions that map from the Boolean cube to the real numbers. 
We equip $\mathcal{L}_{n}$ with the scalar product:
\begin{equation}
    \langle F,G \rangle:=  \underset{x}{\mathbb{E}}[F(x)G(x)],
\end{equation}
where $\underset{x}{\mathbb{E}}[\ast]$ represents the expected value operator.
This scalar product induces the standard $2$-norm 
\begin{equation}
    \langle F,F \rangle=:\lVert F\rVert_{2}^{2}=\underset{x}{\mathbb{E}}[F(x)^{2}]. 
\end{equation}
Similarly, for a generic $p\in (0,+\infty )$, we have $\lVert F\rVert_{p}:=\underset{x}{\mathbb{E}}[F(x)^{p}]^{1/p}$.
Then, consider the noise operator $T_{\rho}$ for $\rho \in [-1,1]$, which acts on $\mathcal{L}_{n}$ in the following way
\begin{equation}
    [T_{\rho}F](x):= \underset{z \sim \frac{1-\rho}{2}}{\mathbb{E}}[F(x+z)],
\end{equation}
where $z$ is distributed according to $\frac{1-\rho}{2}$, i.e., with every bit sampled independently from the others, with a bias $P(x_{i}=1)=\frac{1-\rho}{2}$. 
This choice of labeling of the noise operator allows for the simple relation
\begin{equation} \label{Eq:powerProp}
    T_{\rho^{2}}=T_{\rho}^{2}.
\end{equation}
Further, notice that the operator is symmetric, thus 
\begin{equation} \label{Eq:SimmProp}
  \langle T_{\rho}F,G \rangle = \langle F,T_{\rho}G \rangle=\langle T_{\sqrt{\rho}}F,T_{\sqrt{\rho}}G \rangle,
\end{equation}
where in the second step the property in Eq.~\eqref{Eq:powerProp} was used. 
Most importantly, the noise operator obeys the Bonami-Beckner-Gross hyper-contractive inequality 
\begin{theorem}\cite{Beckner1975, Bonami1970, Gross1975}
For $F\in \mathcal{L}_{n}$, $ 1 \leq p \leq q $ and $\rho^{2} \leq \frac{p-1}{q-1}$ it holds that
\begin{equation}
    \lVert T_{\rho} F \rVert_{q} \leq  \lVert F \rVert_{p}.
\end{equation}
\end{theorem}

\subsection{Proof of Perfect Parallel Repetition for KV}
First, notice that the winning probability of Alice and Bob can be expressed as 
\begin{equation}
    P_{W}:= \underset{\bm x,z \sim \eta}{\mathbb{E}}[\delta_{a(\bm x)+b(\bm x+z),z}].
\end{equation}
Without loss of generality, we assume their strategy $a(\bm x), b(\bm x)$ to be deterministic. 
Then, we define $A(\bm x):= \delta_{\bm x,a(\bm x)}$, and similarly $B(\bm x'):=\delta_{\bm x',b(\bm x')}$.
In other words, $A,B$ equal 1 for exactly one element of each orbit.
Then, we have   
\begin{equation}
  \delta_{a(\bm x)+b(\bm x+z),z}=\sum_{h\in H_{n}^{L}}A(\bm x+h)B(\bm x+z+h).
\end{equation}
Thus, the winning probability $P_{W}$ is given by 
\begin{align}
    P_{W}&=\sum_{h\in H_{n}^{L}}\underset{\bm x,z\sim \eta}{\mathbb{E}}[A(\bm x+h)B(\bm x+z+h)]\\
    & =\abs{H_{n}^{L}}\underset{\bm x,z\sim \eta}{\mathbb{E}}[A(\bm x)B(\bm x+z)],
\end{align}
where we used that the uniform measure is invariant under translations.
We now use the properties of the noise operator detailed in Eqs.~\eqref{Eq:powerProp} and \eqref{Eq:SimmProp} to obtain 
\begin{align}
    & \underset{\bm x,z\sim \eta}{\mathbb{E}}[A(\bm x)B(\bm x+z)]=\underset{\bm x}{\mathbb{E}}[A(\bm x)\underset{z\sim \eta}{\mathbb{E}}[B(\bm x+z)]]\\ 
    & =\langle A,T_{1-2\eta}B \rangle=\langle T_{\sqrt{1-2\eta}}A,T_{\sqrt{1-2\eta}}B \rangle.
\end{align}
Then, using the Cauchy Schwarz inequality with the symmetric conditions on $A,B$, and the Bonami-Beckner-Gross hyper-contractive inequality:
\begin{align}
    &\langle T_{\sqrt{1-2\eta}}A,T_{\sqrt{1-2\eta}}B \rangle\leq \lVert T_{\sqrt{1-2\eta}} A \rVert_{2}^{2}\\
    & \leq \lVert A \rVert_{2-2\eta}^{2}= \underset{\bm x}{\mathbb{E}}[A(\bm x)]^{\frac{1}{1-\eta}}=n^{-\frac{L}{1-\eta}}, \label{eq:finalbound}
\end{align}
where we used that $A(\bm x)=A(\bm x)^{2-2\eta}$ since $A(\bm x)\in \{0,1\}$ and that $\underset{\bm x}{\mathbb{E}}[A(\bm x)]=\underset{\bm x}{\mathbb{E}}[B(\bm x)]=n^{-L}$.
This allows us to write the winning probability as $P_{W}=n^{L}n^{-\frac{L}{1-\eta}}=n^{-L\frac{\eta}{1-\eta}}$, showing that the winning probability is the product of the local winning probabilities, i.e., $P_{W}^{\otimes L}=P_{W}^L$. To achieve the highest score we fix the free parameter $\eta:= \frac{1}{2}-(\log_{2}(n))^{-1}$.  We remark that bound is asymptotically saturable in $n$. 
Indeed, in \cite{BuhrmannRegevScarpadeWolf2011} a classical strategy to asymptotically saturate \eqref{eq:finalbound} for $L=1$ was suggested. 
Thus, following the same classical strategy for each individual input independently saturates asymptotically in $n$ \eqref{eq:finalbound} for any finite $L$. 
For completeness, the strategy for $L=1$ consists of  both parties choosing the bit-string in their input orbit with the highest Hamming weight. 
For generic $L$, the asymptotically optimal strategy consists of both parties choosing $L$ strings of $n$ bits, each obtained by joining together the $L$ bit-strings with the highest Hamming weight of each respective input orbit, as outputs.


\section{Proof for Theorem 4}\label{app:GMNLactivationlemma}
We now prove Lemma~1. 
For this, we consider the following state, which is prepared in a network $\Gamma$,
\begin{align}\label{eq:starfractionstate}
    \sigma(F) \coloneqq F\Phi^{\Gamma}+ (1-F)\tilde\rho,
\end{align}
where $F\coloneqq \Tr(\sigma(F)\Phi^{\Gamma})$. 
Additionally, this implies that $\Tr(\Phi^{\Gamma}\tilde\rho)=0$.
It is important to note, that $\tilde\rho$ is not necessarily positive semidefinite (PSD), i.e., 
\begin{align}
    \sigma(F) - F \Phi^{\Gamma} \nsucceq 0,
\end{align}
unless $|\Phi^{\Gamma}\rangle$ is an eigenvector of $\sigma(F)$.
However, we can transform $\tilde\rho$ to be PSD, by applying the isotropic twirling operation
\begin{align}
    \mathcal{T}[\rho_{AB}]\coloneqq \int  U_A\otimes U_B^* \rho_{AB} U_A^\dagger \otimes {U_B^*}^\dagger dU
\end{align}
in each link of the network, which we denote as $\mathcal{T}^{\otimes N}(\rho)$, which transforms the state in each link into the isotropic state $\rho^\mathrm{iso}\coloneqq F\Phi^+\frac{1-F}{d^2-1}(\mathbbm{1}-\Phi^+)$.
To see this, consider that the isotropic twirling operation transforms a state $\rho$ in the following way \cite{Rains2001}
\begin{align}
    \mathcal{T}[{\rho}] = \Phi^+ \Tr(\Phi^+\rho) + (\id-\Phi^+)\Tr((\id-\Phi^+)\rho).
\end{align}
Additionally, this implies that any entanglement between the two subsystems that are twirled and any other subsystems is removed.
Importantly, the twirling operation is an LOSR transformation, and therefore cannot increase nonlocality.
Now, because $(\mathbbm{1}-\Phi^+)$ is PSD, any tensor products of $\Phi^+$ and $(\mathbbm{1}-\Phi^+)$ will be PSD as well, and therefore $\mathcal{T}^{\otimes N}[\tilde\rho]$ is PSD.
Next, we consider $k$ copies of the state in Eq.~\eqref{eq:starfractionstate} after the twirling, i.e.,
\begin{align}
    \sigma'(F)^{\otimes k}\coloneqq (\mathcal{T}^{\otimes N}[\sigma(F)])^{\otimes k}= F^k {\Phi^{\Gamma}}^{\otimes k}+\ldots.
\end{align}
To compute the Bell violation of $\sigma'(F)^{\otimes k}$ for some Bell game, we notice that the normalized Bell violation 
\begin{align}
    \Gamma(\rho)\coloneqq\frac{S(\rho)}{S_L}
\end{align}
is additive for each term in the decomposition of a state, because Bell inequalities are linear, i.e.,
\begin{align}
    \Gamma(\rho) = \sum_i p_i \underset{\ge 0}{\underbrace{\Gamma(\rho_i)}},
\end{align}
where $p_i\ge 0$, $\sum_i p_i\le 1$, and $\rho_i$ are valid states.
If $\rho_i$ is not PSD or $p_i\ngeq 0$, there may be eigenvectors of $\rho$ with negative eigenvalues, which amount to that eigenvector occurring with negative probability, which can reduce the score $S$ and by extension $\Gamma(\rho)$.
Therefore, because we have ${\Phi^{\Gamma}}^{\otimes m}\otimes\mathcal{T}[\tilde\rho]^{\otimes n}\succeq 0$, we can neglect the score that all terms of this form contribute, and we have 
\begin{align}\label{eq:lowerboundquantum}
    \Gamma(\sigma'(F)^{\otimes k}) \ge F^k \Gamma({\Phi^{\Gamma}}^{\otimes k}).
\end{align}
We now apply Theorem~1, which tells us that $\Gamma(\sigma'(F))$ is GMNL superactivatable, if
\begin{align}\label{eq:quantumvslocalbound}
    \Gamma(\sigma'(F)) > S_L,
\end{align}
where we do not consider a parallel repetition score, because the capacity $c$ of the min-cut of a network obeys $c=1$.
We can now insert Eq.~\eqref{eq:lowerboundquantum} into Eq.~\eqref{eq:quantumvslocalbound}, and
\begin{align}\label{eq:quantumvslocalboundlowerbound}
    F^k\Gamma^\Gamma({\Phi^{\Gamma}}^{\otimes k}) > S_L,
\end{align}
implies that $\sigma'(F)$ is GMNL superactivatable.
Now, we assume that the bipartite game that we use in Theorem~1 is the KV game.
Then, inserting the score for the quantum strategy, that is achieved by playing each instance in each link of the network independently, and inserting the local bound for a single repetition, we get
\begin{align}
    F^k \frac{C'}{\ln^2(d^k)} > \frac{C}{d^k}\quad\implies \quad(Fd)^k>1.
\end{align}
Therefore, if $Fd>1$, our violation diverges, and for some $k\ge k'$ we will achieve a Bell violation.
In conclusion, we have
\begin{align}
    F>\frac{1}{d}\quad\implies \quad \sigma(F)\mathrm{\ is \ GMNL\ superactivatable}.\label{eq:activationthreshold}
\end{align}

\end{document}